\documentclass[conference,onecolumn,12pt]{IEEEtran}
\usepackage[cmex10]{amsmath}
\usepackage{amssymb}
\usepackage{amsthm}
\usepackage{epsfig}
\usepackage{color}

\DeclareMathOperator*{\argmin}{argmin}
\DeclareMathOperator*{\modl}{mod}
\DeclareMathOperator*{\Var}{Var}

\begin{document}
\title{Communicating the Difference of Correlated Gaussian Sources Over a MAC}
\author{\IEEEauthorblockN{Rajiv Soundararajan and Sriram Vishwanath}
\IEEEauthorblockA{Department of Electrical and Computer Engineering, The University of Texas at Austin\\
1 University Station C0803, Austin, TX 78712, USA\\
Email: rajiv.s@mail.utexas.edu and sriram@ece.utexas.edu}}
\maketitle
\begin{abstract}
This paper considers the problem of transmitting the difference of two jointly Gaussian sources over a two-user additive Gaussian noise multiple access channel (MAC). The goal is to recover this difference within an average mean squared error distortion criterion. Each transmitter has access to only one of the two Gaussian sources and is limited by an average power constraint. In this work, a lattice coding scheme that achieves a distortion within a constant of a distortion lower bound is presented if the signal to noise ratio (SNR) is greater than a threshold. Further, uncoded transmission is shown to be worse in performance to lattice coding methods. An alternative lattice coding scheme is presented that can potentially improve on the performance of uncoded transmission. 
\end{abstract}
\section{Introduction}
In this paper, we consider the joint source channel coding problem of transmitting the difference of two positively correlated Gaussians in a distributed fashion over an additive Gaussian noise multiple access channel (MAC). Each transmitter in the MAC has, as its message, one component of the bivariate Gaussian source and its codebook is constrained by a second moment (average power) requirement. We estimate the difference between the two correlated sources while incurring the lowest possible mean squared error at the receiver. The distortion suffered by the difference between the two sources is a function of the power constraints at the two transmitters as well as channel statistics. In general, there is no separation between source and channel coding over MACs, and a joint coding scheme is desired.

There has been significant related work on both the source and channel aspects of this problem. In \cite{Lapidoth2006}, the authors consider the problem of communicating a bivariate Gaussian source over a Gaussian MAC to recover {\em both} components limited by individual distortion constraints. In \cite{Gastpar2007}, the problem of recovering a single Gaussian source through a Gaussian sensor network is considered. Subsequently, the authors also address the problem communicating the sum of independent Gaussian sources over a Gaussian MAC in \cite{Nazer2008}. 

In  the domain of source coding, \cite{Wagner2006} considers and solves the two terminal Gaussian source coding problem while a distributed lattice based coding scheme for reconstructing a linear function of jointly Gaussian sources is developed in \cite{Krithivasan2007}. In \cite{Wagner2008}, an outer bound on the rate region for the distributed compression of linear functions of two Gaussian sources for certain correlations is presented. This bound indicates that existing achievable schemes are suboptimal.

In this work, we present a lattice coding scheme for the distributed transmission of the difference of Gaussians over the MAC.  Note that, for a different setting, lattice codes have been previously considered for joint source-channel coding in \cite{Kochman2008}.  The key contributions of this paper are as follows:
\begin{enumerate}
\item We present a lower bound on the distortion incurred while estimating the difference between the sources over a Gaussian MAC. This lower bound is based on augmenting the receiver with a random variable that induces conditional independence between the two sources and considering a statistically equivalent system of two parallel channels from each of the transmitters to the same receiver. This genie aided bound is  based on the work in \cite{Wagner2006} and \cite{Wagner2008} where the authors determine a lower bound on distortion in a source coding setting.
\item We develop a lattice coding scheme for communicating the difference of the two sources over this channel. The scheme we present for the MAC is similar in spirit to the scheme in \cite{Nazer2008} and is an extension of \cite{Nazer2008} to correlated sources.
\item  We show that our scheme performs ``close'' to the lower bound by showing that the logarithm of the ratio of the distortion achieved to the distortion lower bound is 1 bit if the signal to noise ratio (SNR) is greater than a threshold. We show that the lattice based transmission scheme provides an improvement in distortion over uncoded transmission.
\item Finally, we propose a common dither based lattice coding scheme in which the channel inputs of the two users are correlated (by using the same dither). This correlation can potentially reduce the distortion and can therefore come closer to the lower bound in terms of performance.
\end{enumerate}

The rest of the paper is organized as follows. We develop the system model and notation in Section \ref{sec:sysmod}. We present a lower bound on achievable distortion in Section \ref{sec:obound}. In Section \ref{sec:utran}, we characterize the distortion achieved using an uncoded transmission scheme. In Sections \ref{sec:slattice} and \ref{sec:cdlattice} we describe the scaled lattice and common dither based lattice coding schemes and analyze their performance. Finally, we conclude the paper with Section \ref{sec:conc}. 

\section{System Model and Notation}\label{sec:sysmod}
We briefly explain the notation used in this paper before presenting the system model. We use capitals to denote random variables and boldface capitals to denote matrices. $\mathbb{E}$ is used for expectation of a random variable while we refer to an $n$-length vector as $x^{n}$. Throughout the paper, logarithms used are with respect to base 2 and the square of the 2-norm of an $n$-length vector $x^{n}$ is denoted as
\begin{equation*}
\lVert x^{n}\rVert^{2}_{2}=\sum_{i=1}^{n}(x(i))^2.
\end{equation*}
\\
\begin{figure}[!th]
\centering
\scalebox{0.7}{
\input{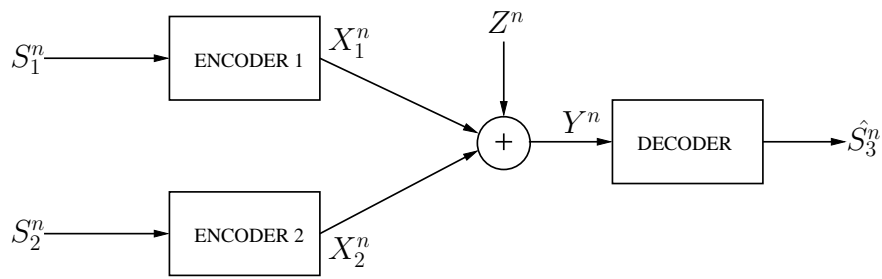}}
\caption{System Model}
\end{figure}
The system model is depicted in Fig. 1. Consider independent and identically distributed (i.i.d) $n$-length sequences of Gaussian random variables, $\{S_{1}(i)\}_{i=1}^{n}$ and $\{S_{2}(i)\}_{i=1}^{n}$. The covariance matrix of $(S_{1}(i),S_{2}(i))$ is given by 
\begin{equation*}
\mathbf{\Sigma} = \left[\begin{array}{ccc} \sigma^{2} & \rho\sigma^{2} \\
\rho\sigma^{2} & \sigma^{2} \end{array}\right]
\end{equation*}
for all $i=1,2,...,n$. Without loss of generality, we assume $\rho > 0$ for the purposes of this paper. Transmitter $k$ in the MAC has a realization of $S_{k}^{n}$ for $k=1,2$. Also the number of source samples observed is equal to the number of channel uses available. Thus, the system has a bandwidth expansion factor of 1. The channel input sequence at each user is a function of the observed source sequence such that a power constraint is satisfied. Mathematically, the channel input $\{X_{k}(i)\}_{i=1}^{n} = f_{k}(\{S_{k}(i)\}_{i=1}^{n})$ for $k=1,2$. The power constraint is expressed as
\begin{equation*}
\frac{1}{n}\sum_{i=1}^{n}\mathbb{E}[(X_{k}(i))^2] \leq P.
\end{equation*}
The noise $\{Z(i)\}_{i=1}^{n}$ is a sequence of i.i.d Gaussian random variables with zero mean and variance $N$. The received signal at time instant $i$ is given by
\begin{equation*}
Y(i) = X_{1}(i)+X_{2}(i)+Z(i).
\end{equation*}
We wish to estimate the sequence of the difference $\{S_{1}(i)-S_{2}(i)\}_{i=1}^{n}$ at the receiver given the received sequence $\{Y(i)\}_{i=1}^{n}$ within a distortion. The distortion metric considered is the time average mean squared error. Let $S_{3}(i)=S_{1}(i)-S_{2}(i)$ and the estimated sequence be $\{\hat{S}_{3}(i)\}_{i=1}^{n}$. The distortion $D$ is defined as 
\begin{equation*}
D = \frac{1}{n}\sum_{i=1}^{n}\mathbb{E}[(S_{3}(i)-\hat{S}_{3}(i))^2].
\end{equation*}
Next, we present a lower bound on $D$.

\section{Lower Bound  on Distortion When Determining the Difference of Jointly Gaussian sources}\label{sec:obound}
We now present a lower bound on the distortion incurred for the distributed transmission of the difference of correlated sources. One of the ideas used in the proof is augmenting the receiver with a random variable that induces conditional independence between $S_{1}$ and $S_{2}$ as presented in \cite{Wagner2008}. We consider the following representation for the Gaussian sources $(S_{1},S_{2})$:
\begin{align*}
&S_{1} = \sqrt{\rho}S + V_{1}\\
&S_{2} = \sqrt{\rho}S + V_{2}.
\end{align*}
where $S$, $V_{1}$ and $V_{2}$ are independent Gaussian random variables with mean zero and variances $\sigma^{2}$, $\sigma^{2}(1-\rho)$ and $\sigma^{2}(1-\rho)$. Note that, by supplying the receiver with the sequence $S^{n}$, the distortion incurred can only decrease. 

\begin{figure}[!th]
\centering
\scalebox{0.7}{
\input{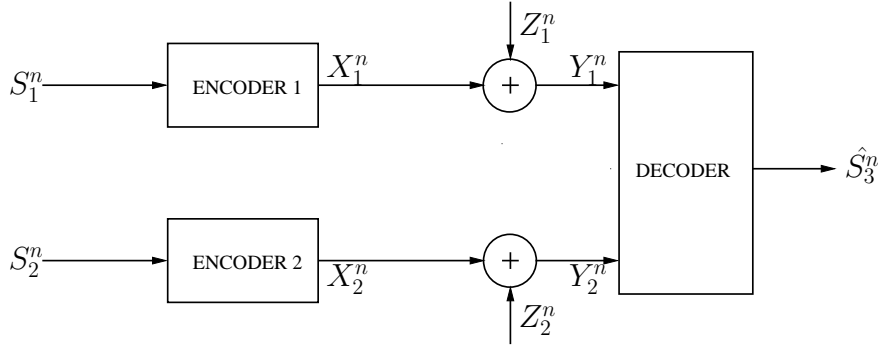}}
\caption{Parallel channels}
\end{figure}

Further, we lower bound the distortion by considering a modified channel setting as shown in Fig. 2. This modified channel is a memoryless Gaussian channel which at time $i$ is represented mathematically as 
\begin{align*}
Y_{1}(i) = X_{1}(i)+Z_{1}(i)\\
Y_{2}(i) = X_{2}(i)+Z_{2}(i)
\end{align*}
where $Z_{1}(i)$ and $Z_{2}(i)$ are Gaussian random variables with mean zero and variance $N/2$, independent of each other and of $X_{1}(i)$ and $X_{2}(i)$. The receiver obtains an estimate of the difference based on the observations of the vector $(Y_{1}^{n},Y_{2}^{n})$. The distortion incurred on this channel is a lower bound on the distortion resulting from the original channel. In the original channel, the output $X_{1}^{n}+X_{2}^{n}+Z^{n}$ is a function of the output of the modified channel, which is the vector $(X_{1}^{n}+Z_{1}^{n},X_{2}^{n}+Z_{2}^{n})$. Note that the output of the original channel (in Fig. 1) and the sum of the outputs of the modified channel (in Fig. 2) are statistically equivalent. 

The distortion incurred in the modified channel with side information $S^n$ at the receiver satisfies:
\begin{align*}
D \geq &\frac{1}{n}\sum_{i=1}^{n}\mathbb{E}[(S_{1}^{n}(i)-S_{2}^{n}(i)-\mathbb{E}[S_{1}^{n}(i)|S^{n},Y_{1}(i),Y_{2}(i)]+\mathbb{E}[S_{2}^{n}(i)|S^{n},Y_{1}(i),Y_{2}(i)])^2] \\
=& \frac{1}{n}\sum_{i=1}^{n}\mathbb{E}[(S_{1}^{n}(i)-\mathbb{E}[S_{1}^{n}(i)|S^{n},Y_{1}(i),Y_{2}(i)])^2]+\mathbb{E}[(S_{2}^{n}(i)-\mathbb{E}[S_{2}^{n}(i)|S^{n},Y_{1}(i),Y_{2}(i)])^2] \\
&-2\mathbb{E}[(S_{1}^{n}(i)-\mathbb{E}[S_{1}^{n}(i)|S^{n},Y_{1}(i),Y_{2}(i)])(S_{2}^{n}(i)-\mathbb{E}[S_{2}^{n}(i)|S^{n},Y_{1}(i),Y_{2}(i)])].
\end{align*}
The following Markov condition
\begin{equation}\label{eqn:markov}
Y_{1}^{n}\leftrightarrow X_{1}^{n}\leftrightarrow S_{1}^{n}\leftrightarrow S^{n}\leftrightarrow S_{2}^{n}\leftrightarrow X_{2}^{n}\leftrightarrow Y_{2}^{n},
\end{equation}
implies that
\begin{align*}
&\mathbb{E}[S_{1}^{n}(i)|S^{n},Y_{1}(i),Y_{2}(i)] = \mathbb{E}[S_{1}^{n}(i)|S^{n},Y_{1}(i)]\nonumber\\
&\mathbb{E}[S_{2}^{n}(i)|S^{n},Y_{1}(i),Y_{2}(i)] = \mathbb{E}[S_{1}^{n}(i)|S^{n},Y_{2}(i)]. 
\end{align*}
Therefore,
\begin{align}\label{eqn:sumdist}
D \geq 
&\frac{1}{n}\sum_{i=1}^{n}\mathbb{E}[(S_{1}^{n}(i)-\mathbb{E}[S_{1}^{n}(i)|S^{n},Y_{1}(i)])^2]+\mathbb{E}[(S_{2}^{n}(i)-\mathbb{E}[S_{2}^{n}(i)|S^{n},Y_{2}(i)])^2] \nonumber\\
&-2\mathbb{E}[(S_{1}^{n}(i)-\mathbb{E}[S_{1}^{n}(i)|S^{n},Y_{1}(i)])(S_{2}^{n}(i)-\mathbb{E}[S_{2}^{n}(i)|S^{n},Y_{2}(i)])].
\end{align}
We observe that
\begin{align}\label{eqn:cond}
&\frac{1}{n}\sum_{i=1}^{n}\mathbb{E}[(S_{1}^{n}(i)-\mathbb{E}[S_{1}^{n}(i)|S^{n},Y_{1}(i)])^2] = \frac{\sigma^{2}(1-\rho)}{1+\frac{2P}{N}} \nonumber\\
&\frac{1}{n}\sum_{i=1}^{n}\mathbb{E}[(S_{2}^{n}(i)-\mathbb{E}[S_{2}^{n}(i)|S^{n},Y_{2}(i)])^2] = \frac{\sigma^{2}(1-\rho)}{1+\frac{2P}{N}}
\end{align}
since these are the average squared error distortions in $S_{1}^{n}$ and $S_{2}^{n}$ when each is transmitted over a point to point Gaussian channel with noise variance $N/2$, power constraint $P$ and conditional variance $\Var(S_{1}|S) = \Var(V_{1}) = \sigma^{2}(1-\rho)$ and $\Var(S_{2}|S) = \Var(V_{2}) = \sigma^{2}(1-\rho)$. 
Now, the following conditional cross correlation,
\begin{align*}
&\mathbb{E}[(S_{1}^{n}(i)-\mathbb{E}[S_{1}^{n}(i)|S^{n},Y_{1}(i)])(S_{2}^{n}(i)-\mathbb{E}[S_{2}^{n}(i)|S^{n},Y_{2}(i)])|S^{n}]\\
=&\mathbb{E}[(S_{1}^{n}(i)-\mathbb{E}[S_{1}^{n}(i)|S^{n},Y_{1}(i)])|S^{n}]\mathbb{E}[(S_{2}^{n}(i)-\mathbb{E}[S_{2}^{n}(i)|S^{n},Y_{2}(i)])|S^{n}]
\end{align*}
due to the Markov condition stated in (\ref{eqn:markov}). But, by tower rule for expectations, we have 
\begin{equation*}
\mathbb{E}[(S_{1}^{n}(i)-\mathbb{E}[S_{1}^{n}(i)|S^{n},Y_{1}(i)])|S^{n}] = 0 \textrm{ a.s},
\end{equation*}
for all $i=1,2,\ldots,n$. Thus  
\begin{align}\label{eqn:cross}
&\mathbb{E}[(S_{1}^{n}(i)-\mathbb{E}[S_{1}^{n}(i)|S^{n},Y_{1}(i),Y_{2}(i)])(S_{2}^{n}(i)-\mathbb{E}[S_{2}^{n}(i)|S^{n},Y_{1}(i),Y_{2}(i)])|S^{n}] = 0 \textrm{ a.s.} \nonumber\\
\Rightarrow &\mathbb{E}[(S_{1}^{n}(i)-\mathbb{E}[S_{1}^{n}(i)|S^{n},Y_{1}(i),Y_{2}(i)])(S_{2}^{n}(i)-\mathbb{E}[S_{2}^{n}(i)|S^{n},Y_{1}(i),Y_{2}(i)])] = 0,
\end{align}
 By combining (\ref{eqn:sumdist}), (\ref{eqn:cond}) and (\ref{eqn:cross}), we get 
\begin{equation*}
D_{bound} = \frac{2\sigma^{2}(1-\rho)}{1+\frac{2P}{N}}.
\end{equation*}

In the following sections, we discuss the performance of various achievable schemes relative to this distortion bound. 
\section{Uncoded Transmission}\label{sec:utran}
In this section, we compute the distortion resulting from  uncoded transmission  in order to communicate the difference $\{S_{1}(i)-S_{2}(i)\}_{i=1}^{n}$. In this setting, Transmitter 1 sends a scaled version of the source  $\sqrt{\frac{P}{\sigma^{2}}}S_{1}(i)$ at time instant $i$ and Transmitter 2 sends $-\sqrt{\frac{P}{\sigma^{2}}}S_{2}(i)$ at time instant $i$. The scaling is chosen such that both users satisfy their respective power constraints. The  received sequence is given by
\begin{equation*}
Y(i) = \sqrt{\frac{P}{\sigma^{2}}}(S_1(i)-S_2(i))+Z(i).\\
\end{equation*}
The receiver determines the minimum mean squared error (MMSE) estimate of the difference $S_1(i)-S_2(i)$ based on the received signal $Y(i)$. The distortion resulting from this process can be calculated as
\begin{align*}
D = &\frac{1}{n}\mathbb{E}[\lVert S_{3}^{n}-\mathbb{E}[S_{3}^{n}|Y^{n}]\rVert^{2}_{2}]\\
=&2\sigma^{2}(1-\rho) - \frac{(2\sqrt{P\sigma^{2}}(1-\rho))^2}{2P(1-\rho)+N}\\
=&\frac{2\sigma^{2}(1-\rho)}{1+\frac{2P(1-\rho)}{N}}.
\end{align*}
Note that the distortion resulting from by uncoded transmission does not meet the lower bound for any $\rho > 0$. In the next two sections, we describe lattice based coding schemes which  perform better than uncoded transmission (thus resulting in a lower distortion). 

\section{Lattice Coding Scheme}\label{sec:slattice}
We now describe a scaling based lattice scheme to communicate the difference of the two sources. An implicit assumption that we make in the study and design of lattice schemes is that $P\leq\sigma^{2}$. In effect, for $P > \sigma^{2}$,  the lattice quantization scheme presented below  reduces to uncoded transmission. We briefly review some features of lattice codes and quantizers before we present the scheme. A lattice of dimension $n$ is defined as the set
\begin{equation*}
\Lambda = \{x = z\mathbf{G}: z\in \mathbb{Z}^{n}\}
\end{equation*}
where $\mathbf{G}\in \mathbb{Z}^{n\times n}$ is known as the generator matrix and $\mathbb{Z}$ is the set of all integers. The quantized value of $x\in \mathbb{R}^{n}$, $Q(x)=\argmin_{r\in \Lambda} \lVert x-r\rVert$. The second moment of a lattice is defined as $\sigma^{2}(\Lambda)=\frac{\int_{\nu}\lVert x\rVert^{2}dx}{\int_{\nu}dx}$. The fundamental Voronoi region of $\Lambda$ is defined as $\nu_{0}=\{x\in\mathbb{R}^{n}: Q(x)=0\}$. Further, we use the notation $x \modl \Lambda = x - Q(x)$. 

The lattice coding scheme described below is similar in nature to the lattice coding scheme used in \cite{Nazer2008} for joint source channel coding of the sum of independent Gaussian sources. Consider $\Lambda$, a lattice of dimension $n$ with second moment $\sigma^{2}(\Lambda)=P$. We choose the same lattice $\Lambda$ at both the users such that it is good for both source and channel coding. The proof of existence of such a lattice and its construction are detailed in \cite{Erez2005}. Let $U_{1}^{n}$ and $U_{2}^{n}$ be independent dithers (independent of themselves and independent of the sources) which are uniformly distributed over the fundamental Voronoi region $\nu_{0}$ and known at the receiver. The $n$-length channel input at each transmitter is 
\begin{align*}
X_{1}^{n} &=(\gamma S_{1}^{n}-U_{1}^{n}) \modl \Lambda \\
X_{2}^{n} &=(-\gamma S_{2}^{n}-U_{2}^{n}) \modl \Lambda
\end{align*}
where $\gamma$ is a scalar which is chosen later. 

The signal at the receiver is given by 
\begin{equation*}
Y^{n} = X_{1}^{n}+X_{2}^{n}+Z^{n}. 
\end{equation*}
The decoder performs the following operations to estimate the difference.
\begin{align*}
Y^{n}_{1}=&[\alpha Y^{n}+U_{1}^{n}+U_{2}^{n}]\modl \Lambda\\
=&[\alpha(X_{1}^{n}+X_{2}^{n}+Z^{n})+U_{1}^{n}+U_{2}^{n}]\modl \Lambda \\
=&[\gamma(S_{1}^{n}-S_{2}^{n})+(\alpha-1)((\gamma S_{1}^{n}-U_{1}^{n})\modl \Lambda \\
&+(-\gamma S_{2}^{n}- U_{2}^{n})\modl \Lambda)+\alpha Z^{n}]\modl \Lambda\\
=&[\gamma(S_{1}^{n}-S_{2}^{n})+Z^{n}_{1}]\modl \Lambda,
\end{align*}
where 
\begin{equation*}
Z_{1}^{n}=(\alpha-1)((\gamma S_{1}^{n}-U_{1}^{n})\modl \Lambda+(-\gamma S_{2}^{n}- U_{2}^{n})\modl \Lambda)+\alpha Z^{n}
\end{equation*}
is the effective noise. Note that each term in the effective noise is independent of the source since the dither is chosen uniformly in the fundamental Voronoi region and independent of the sources \cite{Erez2004} and the original noise $Z^{n}$ is also independent of the sources. By choosing $\alpha=\frac{2P}{2P+N}$, the MMSE coefficient, we reduce the variance of the effective noise to $\frac{2PN}{2P+N}$. Since $\Lambda$ is chosen to be a good channel lattice, if
\begin{equation}\label{eqn:corrdec}
\gamma^{2}2\sigma^{2}(1-\rho)+\frac{2PN}{2P+N}\leq P,
\end{equation}
we know from \cite{Poltyrev1994} that we can decode correctly and 
\begin{equation*}
[\gamma(S_{1}^{n}-S_{2}^{n})+Z_{1}^{n}]\modl \Lambda = \gamma(S_{1}^{n}-S_{2}^{n})+Z_{1}^{n}.
\end{equation*}
Therefore, if $\frac{P}{N} > \frac{1}{2}$, we choose $\gamma$ satisfying (\ref{eqn:corrdec}) with equality. Mathematically, $\gamma$ satisfies 
\begin{align}\label{eqn:gamma}
& \gamma^{2}2\sigma^{2}(1-\rho)+\frac{2PN}{2P+N} = P \nonumber\\
\Rightarrow & \gamma^{2}2\sigma^{2}(1-\rho)\frac{2P+N}{2PN} + 1  = \frac{2P+N}{2N}. %= SNR + \frac{1}{2}
\end{align}
Under the assumption of correct decoding, we have
\begin{equation*}
Y_{1}^{n} = \gamma(S_{1}^{n}-S_{2}^{n})+Z_{1}^{n}.
\end{equation*}
We now multiply the received signal by $\frac{1-K}{\gamma}$ where $K=\frac{2PN}{2PN+2\sigma^{2}(1-\rho)\gamma^{2}(2P+N)}$, to obtain 
\begin{align*}
\hat{S}_{3}^{n} &= \frac{1-K}{\gamma}(\gamma S_{3}^{n}+Z_{1}^{n})\\
&= S_{3}^{n} - KS_{3}^{n}+\frac{1-K}{\gamma}Z_{1}^{n}.
\end{align*}
The average distortion that is achieved is simply the time average of the expectation of the two norm of $\frac{1-K}{\gamma}Z_{1}^{n}- KS_{3}^{n}$, which can be calculated as
\begin{align*}
D_{lattice} &= \frac{\mathbb{E}[\lVert \frac{1-K}{\gamma}Z_{1}^{n}- KS_{3}^{n} \rVert^{2}]}{n}\\
&= \frac{2\sigma^{2}(1-\rho)}{1+\frac{2\sigma^{2}(1-\rho)\gamma^{2}(2P+N)}{2PN}}\\
&= \frac{2\sigma^{2}(1-\rho)}{\frac{P}{N}+\frac{1}{2}}. 
\end{align*}
where the last equality follows from (\ref{eqn:gamma}). 

The lattice based coding scheme developed above is close to the distortion bound presented in Section \ref{sec:obound} in the sense that the logarithm of the ratio of the distortion bound to distortion resulting from the lattice scheme is one bit for any $SNR>\frac{1}{2}$. This is because 
\begin{equation*}
\log\frac{D_{lattice}}{D_{bound}} = \log \frac{2\sigma^{2}(1-\rho)}{\frac{P}{N}+\frac{1}{2}}\frac{1+\frac{2P}{N}}{2\sigma^{2}(1-\rho)} = \log 2 = 1.
\end{equation*}
The SNR condition is necessary for the existence of the above lattice scheme as discussed earlier. 

\section{Common Dither based Lattice Coding Scheme}\label{sec:cdlattice}
We now propose an alternative lattice coding scheme based on using a common dither at both the terminals. Let $U^{n}$ be the common dither at both the terminals and the rest of the parameters of the lattice code are the same as in the previous section. The channel input at each user is given by
\begin{align*}
X_{1}^{n} &=(S_{1}^{n}-U^{n}) \modl \Lambda \\
X_{2}^{n} &=-(S_{2}^{n}-U^{n}) \modl \Lambda.
\end{align*}
We know that $X_{k}^{n}$ is independent of $S_{k}^{n}$ for $k=1,2$ and is uniformly distributed over the fundamental Voronoi region of the lattice $\Lambda$ \cite{Erez2004}. However $X_{1}^{n}$ and $X_{2}^{n}$ are no longer independent. Let $\rho'$ denote the correlation coefficient between $X_{1}^{n}$ and $X_{2}^{n}$. In this scheme, we perform the same sequence of operations at the receiver as in the previous lattice based scheme. Thus we obtain 
\begin{equation*}
Y_{1}^{n}=[S_{1}^{n}-S_{2}^{n}+Z_{1}^{n}]\modl \Lambda,
\end{equation*}
where 
\begin{equation*}
Z_{1}^{n}=(\alpha-1)((S_{1}^{n}-U^{n})\modl \Lambda-(S_{2}^{n}-U^{n})\modl \Lambda)+\alpha Z^{n}
\end{equation*}
is the effective noise. By choosing $\alpha = \frac{2P(1+\rho')}{2P(1+\rho')+N}$, the variance of $Z_{1}^{n}$ can be reduced to  $\frac{2P(1+\rho')N}{2P(1+\rho')+N}$. Again, as before, the effective noise term is independent of $S_{1}^{n}-S_{2}^{n}$. Moreover, $[S_{1}^{n}-S_{2}^{n}+Z_{1}^{n}]\modl \Lambda = S_{1}^{n}-S_{2}^{n}+Z_{1}^{n}$ if 
\begin{equation*}
2\sigma^{2}(1-\rho)+\frac{2P(1+\rho')N}{2P(1+\rho')+N}\leq P. 
\end{equation*}
Thus we will be able to decode correctly for all $P$, $N$, $\rho$ and $\rho'$ satisfying the above equation. Multiplying the signal $Y_{1}^{n}$ by $1-K$ where $K = \frac{2P(1+\rho')N}{2P(1+\rho')N+2\sigma^{2}(1-\rho)(2P(1+\rho')+N)}$, the net distortion can be calculated similarly as 
\begin{align*}
D &= \frac{\mathbb{E}[\lVert (1-K)Z_{1}^{n}- KS_{3}^{n} \rVert^{2}]}{n}\\
&= \frac{2\sigma^{2}(1-\rho)}{1+\frac{2\sigma^{2}(1-\rho)(2P(1+\rho')+N)}{2P(1+\rho')N}} 
\end{align*}
In general, the distortion resulting from the common dither based scheme is better than that resulting from the independent dither based scheme. This improvement depends on $\rho'$, the correlation between the channel inputs. Characterizing $\rho'$ is in general a non-trivial task as it depends on both source and channel parameters, and is therefore left uncharacterized in this paper. 
\section{Conclusion}\label{sec:conc}
We present two lattice coding schemes for the distributed source channel communication of the difference of two jointly Gaussian sources. In the scaling based lattice coding scheme, we show that we can find the scaling parameter $\gamma$ to achieve a distortion very close to the lower bound on the distortion if $SNR>\frac{1}{2}$. Future work includes exploring lattice based schemes to compute more general linear functions of correlated Gaussian sources over a MAC. 
\section{Acknowledgment}
The authors thank Aaron Wagner and Ram Zamir for their helpful comments.


\begin{thebibliography}{10}
\providecommand{\url}[1]{#1}
\csname url@samestyle\endcsname
\providecommand{\newblock}{\relax}
\providecommand{\bibinfo}[2]{#2}
\providecommand{\BIBentrySTDinterwordspacing}{\spaceskip=0pt\relax}
\providecommand{\BIBentryALTinterwordstretchfactor}{4}
\providecommand{\BIBentryALTinterwordspacing}{\spaceskip=\fontdimen2\font plus
\BIBentryALTinterwordstretchfactor\fontdimen3\font minus
  \fontdimen4\font\relax}
\providecommand{\BIBforeignlanguage}[2]{{%
\expandafter\ifx\csname l@#1\endcsname\relax
\typeout{** WARNING: IEEEtran.bst: No hyphenation pattern has been}%
\typeout{** loaded for the language `#1'. Using the pattern for}%
\typeout{** the default language instead.}%
\else
\language=\csname l@#1\endcsname
\fi
#2}}
\providecommand{\BIBdecl}{\relax}
\BIBdecl

\bibitem{Lapidoth2006}
A.~Lapidoth and S.~Tinguely, ``Sending a bi-variate {G}aussian source over a
  {G}aussian {MAC},'' in \emph{Proc. IEEE Int Symp Info Theory}, Seattle, WA
  2006.

\bibitem{Gastpar2007}
M.~Gastpar, ``Uncoded transmission is exactly optimal for a simple {G}aussian
  sensor network,'' in \emph{Proc. 2007 ITA Workshop}, San Diego, CA 2007.

\bibitem{Nazer2008}
B.~Nazer and M.~Gastpar, ``Strcutured {R}andom {C}odes and {S}ensor {N}etwork
  {C}oding {T}heorems,'' in \emph{Proceedings of the 20th Biennial
  International Zurich Seminar on Commununication (IZS 2008)}, Zurich,
  Switzerland, 2008.

\bibitem{Wagner2006}
A.~Wagner, S.~Tavildar, and P.~Viswanath, ``Rate region of the {Q}uadratic
  {G}aussian {T}wo-{E}ncoder {S}ource-{C}oding {P}roblem,'' \emph{{IEEE} Trans.
  Inf. Theory}, 2008, submitted for publication. Preprint available at
  http://arxiv.org/abs/cs/0510095.

\bibitem{Krithivasan2007}
D.~Krithivasan and S.~Pradhan, ``Lattices for distributed source coding:
  {J}ointly {G}aussian sources and reconstruction of a linear function,''
  \emph{{IEEE} Trans. Inf. Theory}, 2007, submitted for publication. Preprint
  available at http://arxiv.org/abs/0707.3461.

\bibitem{Wagner2008}
A.~Wagner, ``An outer bound for distributed compression of linear functions,''
  in \emph{42nd Annual Conference on Information Sciences and Systems (CISS)},
  Princeton, NJ 2008.

\bibitem{Kochman2008}
Y.~Kochman and R.~Zamir, ``Joint {W}yner-{Z}iv/{D}irty-{P}aper {C}oding by
  {M}odulo-{L}attice {M}odulation,'' \emph{{IEEE} Trans. Inf. Theory}, 2008,
  submitted for publication. Preprint available at
  http://www.eng.tau.ac.il/~zamir/publications.html.

\bibitem{Erez2005}
U.~Erez, S.~Litsyn, and R.~Zamir, ``Lattices which are good for (almost)
  everything,'' \emph{{IEEE} Trans. Inf. Theory}, vol.~51, pp. 3401--3416, Oct.
  2005.

\bibitem{Erez2004}
U.~Erez and R.~Zamir, ``Achieving $\frac{1}{2}\log(1+{SNR})$ on the {AWGN}
  channel with lattice encoding and decoding,'' \emph{{IEEE} Trans. Inf.
  Theory}, vol.~50, pp. 2293--2314, Oct. 2004.

\bibitem{Poltyrev1994}
G.~Poltyrev, ``On coding without restructions for the {AWGN} channel,''
  \emph{{IEEE} Trans. Inf. Theory}, vol.~40, pp. 409--417, Mar. 1994.

\end{thebibliography}
\end{document}